\documentclass[twocolumn]{revtex4}

\usepackage{graphicx}
\usepackage{amsmath}
\usepackage{epsfig}
\usepackage{color}

\begin{document}

\title{Lyapunov exponent of the random frequency oscillator: 
       cumulant expansion approach}

\author{C. Anteneodo$^1$ and R. O. Vallejos$^2$}
\address{
$^1$ Department of Physics, PUC-Rio and 
National Institute of Science and Technology for Complex Systems, 
Rua Marqu\^es de S\~ao Vicente 225, G\'avea, CEP 22453-900 RJ, 
Rio de Janeiro, Brazil \\
$^2$ Centro Brasileiro de Pesquisas Fisicas, 
Rua Dr. Xavier Sigaud 150, Rio de Janeiro, Brazil
}



\begin{abstract}
We consider a one-dimensional harmonic oscillator with a 
random frequency, focusing on
both the standard and the generalized 
Lyapunov exponents, 
$\lambda$ and $\lambda^\star$ respectively.
We discuss the numerical difficulties that arise in the 
numerical calculation of $\lambda^\star$ in the case of 
strong intermittency.
When the frequency corresponds to a 
Ornstein-Uhlenbeck process,
we compute analytically $\lambda^\star$ by using
a cumulant expansion including up to the fourth order. 
Connections with the problem of 
finding an analytical estimate for the largest
Lyapunov exponent of a many-body
system with smooth interactions are discussed.
\end{abstract}
 
\maketitle

\section{Introduction}

The theory of Lyapunov exponents of hard-ball systems 
has a long history. 
It started with the pioneering work of Krylov 
\cite{krylov,ma}, 
was rigorously developed by Sinai \cite{sinai} and
collaborators, and
completed (to some extent) by van Beijeren, Dorfman and 
co-workers \cite{vanb1,vanb2,vanzon,kruis,dorfman}. 
The analytical calculation of, e.g., the largest 
Lyapunov exponent of a dilute rigid-sphere gas, is 
based on the fact that the dynamics consists of  
free rectilinear motions interrupted by instantaneous 
elastic collisions \cite{vanzon};
the expressions so-obtained agree quantitatively
with the numerical experiments 
\cite{vanzon,dellago1,dellago2}. 

The case of a dilute gas with finite-range interactions 
can be handled in close analogy with the rigid-sphere 
problem: 
though the collisions are not trivial any more, the
dynamics is still ruled by occasional pairwise 
encounters \cite{vanzon,kimball,elyutin}. 
However, when one considers long-range interactions
(or short-range interactions and high densities),
the theoretical approach must be substantially
modified. 

In the general case we must deal with the full system of 
coupled differential equations that govern the evolution 
of multidimensional tangent vectors
$\eta(t)$. 
Consider for instance a gas of $N$ particles in three
dimensions described by the Hamiltonian
\begin{equation}
\label{ham}
{\cal H} = 
\sum_{i=1}^{3N} \frac{ p^2_i}{2m} + {\cal V}(q_1,\ldots,q_{3N}),
\end{equation}
where $q_i$ and $p_i$, are conjugate position-momentum 
coordinates. 
Assuming $m=1$, tangent vectors evolve according to
\begin{equation}
\label{tangent}
\dot \eta = 
\left( 
\begin{matrix}
    0          &    1   \cr
 -{\bf V}(t)   &    0        
\end{matrix}
 \right)\, \eta  \; 
\end{equation}
(dot meaning time derivative), 
where $\bf V$ is the Hessian matrix of the potential ${\cal V}$, 
namely
\begin{equation} 
\label{V}
V_{ij} =  
\frac{\partial^2{\cal V}}{\partial q_i \partial q_j} \;.
\end{equation}
The Hessian depends explicitly on time because it is calculated
along a reference trajectory $q(t)$.
Once initial conditions $x_0=(q_0,p_0)$ and $\eta_0$ have been 
specified,
one can find $\eta(t)$ from Eq.~(\ref{tangent}).
Then the Lyapunov exponent $\lambda$ is obtained by calculating 
the limit \cite{benettin}
\begin{equation}
\label{liap}
\lambda = 
\lim_{t \to \infty} \frac{1}{t} \ln | \eta (t; x_0,\eta_0)| \;.
\end{equation}
Assuming ergodicity on the energy-shell,
$\lambda$ becomes independent of initial conditions
$x_0$, which can then be chosen randomly according to the 
microcanonical distribution. 
There will also be no dependence on initial tangent vectors,
because if $\eta_0$ is also chosen randomly, it will have a 
non-zero component along the most expanding direction.
It is this average over $x_0$ and $\eta_0$
that permits to treat equations (\ref{tangent})
formally as a system of stochastic differential equations 
\cite{vankampen}.
Moreover, if the dynamics can be thought of as free motion
plus weak interactions, then perturbative techniques, like
the cumulant expansion \cite{vankampen,kubo,fox}, can be 
invoked.
So, the theory attempts to calculate the average
\begin{equation}
\label{liapav}
\lambda = 
\lim_{t \to \infty} 
\frac{1}{t} 
\langle 
\ln | \eta (t; x_0,\eta_0)| 
\rangle \;.
\end{equation}
However, in practice, it is much simpler to develop an estimate for 
the generalized Lyapunov exponent \cite{benzi,castiglione}
\begin{equation}
\label{liapstarav}
\lambda^\star = 
\lim_{t \to \infty} 
\frac{1}{2t} 
\ln \langle 
 | \eta (t; x_0,\eta_0)|^2 
\rangle \;.
\end{equation}
This is essentially the approach followed by 
Barnett et al \cite{barnett1,barnett2,barnett3}, 
Pettini et al \cite{pettini1,pettini2,pettini3},
and the present authors \cite{av1,av2,av3}.
In situations of weak intermittency both exponents are expected 
to be close. If one wishes to use a theoretically calculated
$\lambda^\star$ as an approximation to $\lambda$, then a numerical 
check must be done first to verify that both exponents coincide. 
The cumulant expansion to be discussed below offers an 
analytical expression for $\lambda$ via the replica trick.
Note, however, that the difficulties involved in such a calculation 
are much greater than those we shall face when dealing with 
$\lambda^\star$. 

Though there are some differences among the 
formulations of the three just-mentioned groups, 
it may be said that the main theoretical conclusion 
extracted from that body of work is: if one combines 
the cumulant expansion with some kind of isotropy 
approximation (which may be fully justified in some
cases), the original problem of $6N$ differential 
equations can be reduced to a system of only two equations 
for a ``representative" single degree of freedom:
\begin{equation}
\label{tangent2}
\left( \begin{matrix}
    \dot \eta_1   \cr
    \dot \eta_2           
\end{matrix} \right)
 = 
\left( \begin{matrix}
    0          &    1   \cr
 -\kappa(t)    &    0        
\end{matrix} \right)  
\left( \begin{matrix}
     \eta_1   \cr
     \eta_2           
\end{matrix} \right)\; .
\end{equation}
In this kind of mean-field approximation, the ``curvature" 
$\kappa(t)$ is a scalar stochastic  process, whose cumulants 
can be related to the (operator) cumulants of the Hessian 
${\bf V}(t)$ (see, e.g., \cite{av1}).

The comparison of theoretical results obtained with the
cumulant approach versus numerical simulations has met mixed success.
The agreement is very good for a many-particle system with
bounded weak interactions \cite{av2,av3} and for the 
Fermi-Pasta-Ulam system \cite{pettini3}. 
However, the results for the 1d-XY model \cite{pettini3}, 
for a dense one-component plasma \cite{barnett1,comment},
and for a dilute Lennard-Jones gas \cite{av4} are not
so satisfactory.

The purpose of this paper is to investigate the limits of
validity of the cumulant approach for 
the Lyapunov exponent of a many particle system.
We choose as a starting point the simplified mean-field 
setting (\ref{tangent2}) and consider two possibilities
for $\kappa(t)$.
It has been argued \cite{pettini3} that, for typical chaotic
many-body systems, $\kappa(t)$ should be close to Gaussian
white noise; this is the first
case we shall consider.
In the white-noise case the second-order expansion for 
$\lambda^\star$ is exact, thus this case is ideally 
suited for analyzing the difficulties that appear 
in the {\em numerical} calculation of $\lambda^\star$.
Next, we keep the Gaussian and Markov properties but 
allow for finite correlation times, leading to the 
Ornstein-Uhlenbeck process;
in this case we calculate the fourth cumulant 
contribution to $\lambda^\star$.  
Though it will not be considered here, we also mention the 
interesting situation of $\kappa(t)$ being a Poisson process, 
which appears to be the appropriate choice for modeling the 
tangent-vector dynamics in a dilute 
gas with short-range interactions.

\section{Cumulant expansion for the Kubo oscillator}

Formally, Eq.~(\ref{tangent2}) describes a harmonic 
oscillator with a random frequency $\omega$ such that
$\omega^2=\kappa$ (Kubo oscillator). 
It is worth generalizing this model a bit to account for 
the possibility of damping, i.e., we shall consider an 
oscillator described by the dynamical equation 
\begin{equation}
\ddot q + \alpha \, \dot q + \kappa \, q = 0 \, .
\label{kuboosc}
\end{equation}
Setting $\alpha=0$, $q=\eta_1$, $p=\dot q =\eta_2$,
recovers (\ref{tangent2}).

Some analytical results for the Lyapunov exponent of the
Kubo oscillator (\ref{kuboosc}) can be found in the 
literature (see, e.g., \cite{mallick,leprovost}).
Here we shall restrict ourselves to the analytical 
calculation of the generalized exponent $\lambda^\star$. 
For this purpose we must consider the dynamics of second moments:
\begin{equation}  \label{B}
\frac{d}{dt}
\left(
\begin{array}{c}
q^2 \cr
p^2 \cr
qp
\end{array}
\right)
=
\left(
\begin{array}{ccc}
  0 &  0                & 2                \cr
  0 & -2\alpha          & -2\kappa         \cr
-\kappa &  1                & -\alpha  
\end{array}
\right)
\left(
\begin{array}{c}
q^2 \cr
p^2 \cr
qp
\end{array}
\right) 
\equiv {\bf B}(t)
\left(
\begin{array}{c}
q^2 \cr
p^2 \cr
qp
\end{array}
\right) 
\, .
\end{equation}
Let us think that, in principle, both parameters $\alpha$ and
$\kappa$ are stationary stochastic processes.
If fluctuations are small enough (in a sense that will be 
discussed later), one can obtain the average of the 
second-moment vector using the first terms of the cumulant
expansion, which works as follows \cite{vankampen}. 
First we split the stochastic matrix as an average plus
fluctuations:
\begin{equation} 
{\bf B}(t)={\bf B_0}+{\bf B_1}(t)  \, .
\end{equation}
For long times one has:
\begin{equation}
\frac{d}{dt}
\left \langle
\left(
\begin{array}{c}
q^2 \cr
p^2 \cr
qp
\end{array}
\right)
\right \rangle
={\bf K}
\left \langle
\left(
\begin{array}{c}
q^2 \cr
p^2 \cr
qp
\end{array}
\right) 
\right \rangle \, ,
\end{equation}
where ${\bf K}$ is the $3 \times 3$ matrix given by the 
operator cumulant expansion \cite{vankampen}
\begin{equation} 
\mathbf{K}= 
\mathbf{B}_0 + 
\int_0^\infty 
\left \langle 
\mathbf{B}_1(\tau) 
\, e^{\mathbf{B}_0 \tau} \, 
\mathbf{B}_1(0) 
\right \rangle 
e^{-\mathbf{B}_0 \tau}
d\tau + \ldots \, .
\label{cum}
\end{equation}
Dots stand for third and higher cumulants (some explicit
expressions can be found in \cite{fox}). 
The exponent $\lambda^\star$ is related to the eigenvalue
of ${\bf K}$ that has the largest real part: 
\begin{equation} \label{lambdastar}
\lambda^\star=
\frac{1}{2} 
\max \,
\Re \, 
\left\{ k_1,k_2,k_3 \right\} \, ,
\end{equation}
with $k_i$ the eigenvalues of $\mathbf{K}$.

\section{Gaussian white noise}

When the entries of the fluctuation matrix $\bf B_1$ are 
Gaussian white noise (and only in this case \cite{fox})
the cumulant expansion stops at the second order, i.e., 
Eq.~(\ref{cum}) is exact (without the ellipsis). 
This is the case we consider now.

\subsection{Random frequency}

\begin{figure}[h] 
\begin{center}
\includegraphics*[bb=120 430 520 700, width=0.45\textwidth]{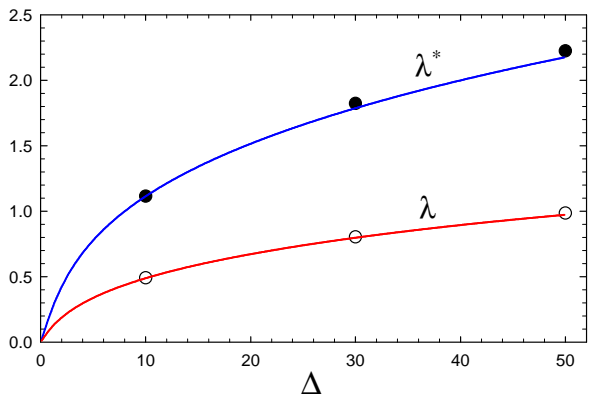}
\end{center}
\caption{\label{fig:RF0} Harmonic oscillator with random frequency.
Shown are the Lyapunov exponents 
$\lambda$ and $\lambda^\star$ as a function of the noise 
strength $\Delta$.
Solid lines correspond to the theoretical 
expressions given by inserting (\ref{K_RF}) into (\ref{lambdastar}), 
and in
Ref.~\cite{mallick} for $\lambda^\star$ and $\lambda$, respectively.
Symbols are the results of numerical simulations 
(averaged over $10^4$ trajectories). 
We chose $\alpha=0$ and $\kappa_0=1$. }
\end{figure}

Let us first study the situation where the damping
$\alpha$ is a constant and
\begin{equation}  \label{kappa}
\kappa(t)=\kappa_0 + \xi(t)   \, ,
\end{equation}
where $\xi(t)$ is a zero-mean Gaussian white noise.
Its correlation function reads
\begin{equation}  
\langle \xi(t) \, \xi(t') \rangle = 
\Delta \, \delta(t-t')  \, .
\end{equation}
With these definitions one has
\begin{equation}  \label{B_RF} 
{\bf B}=
\left(
\begin{array}{ccc}
         0 &  0       &  2            \cr
         0 & -2\alpha & -2 \kappa_0   \cr
 -\kappa_0 &  1       &   -\alpha
\end{array}
\right)
+
\xi(t)
\left(
\begin{array}{ccc}
   0 &  0 & 0  \cr
   0 &  0 & 2  \cr
   1 &  0 & 0 
\end{array}
\right) \, .
\end{equation}
After substitution into Eq.~(\ref{cum}) we readily obtain
\begin{equation} \label{K_RF}
{\bf K}=
\left(
\begin{array}{ccc}
         0 &  0       &  2            \cr
    \Delta & -2\alpha & -2 \kappa_0   \cr
 -\kappa_0 &  1       &   -\alpha
\end{array}
\right) \, .
\end{equation}
The generalized exponent $\lambda^\star$ can now be calculated
from Eq.~(\ref{lambdastar}).
A closed expression for the standard Lyapunov exponent can be 
found in the literature \cite{mallick}.
As an example, Fig.~\ref{fig:RF0} displays  analytical results
for both exponents.
We also show the outcomes of numerical simulations.
Given that the theoretical expressions are exact, 
Fig.~\ref{fig:RF0} constitutes a test for our numerical 
calculations.
Numerical details, including a discussion about the difficulties
found in the calculation of $\lambda^\star$, will be
presented in Sec.~\ref{sec:diff}.
%

\subsection{Random damping}

\begin{figure}[h]
\begin{center}
\includegraphics*[bb=110 390 500 660, width=0.45\textwidth]{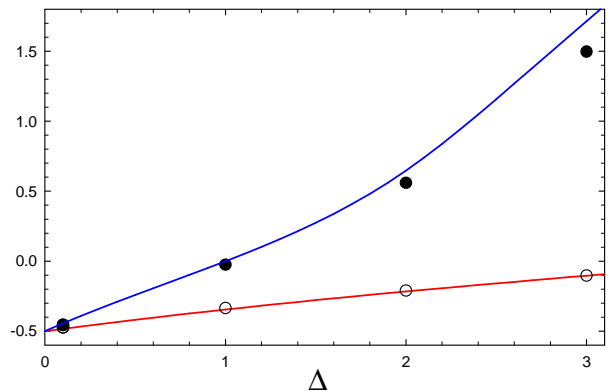}
\end{center}
\caption{ \label{fig:RD0}  Harmonic oscillator with random damping.
Shown are the Lyapunov exponents 
$\lambda$ and $\lambda^\star$
as a function of the
noise strength $\Delta$.
Solid lines correspond to the theoretical expressions given by 
(\ref{lambdastar},\ref{K_RD}) and
Ref.~\cite{leprovost}, for $\lambda^\star$ and $\lambda$, 
respectively.
Symbols are the results of numerical simulations 
(averaged over $10^4$ trajectories). 
We chose $\kappa=1$ and $\alpha_0=1$. 
}
\end{figure}

Now we consider an harmonic oscillator with constant frequency  
but in an environment with 
fluctuating damping coefficient   
\begin{equation}  
\alpha(t)=\alpha_0 + \xi(t)   \, ,
\end{equation}
where $\xi(t)$ is also in this case a zero-mean Gaussian white noise. 
The corresponding stochastic differential equation (\ref{kuboosc}) 
will be taken in 
Stratonovich sense. 
Therefore, the matrix $\mathbf{B}$ in Eq.~(\ref{B}) can be 
decomposed as
\begin{equation} 
{\bf B}=
\left(
\begin{array}{ccc}
         0 &  0       &  2            \cr
         0 & -2\alpha_0 & -2 \kappa   \cr
 -\kappa &  1       &   -\alpha_0
\end{array}
\right)
+
\xi(t)
\left(
\begin{array}{ccc}
   0 &  0 & 0  \cr
   0 &  2 & 0  \cr
   0 &  0 & 1 
\end{array}
\right) \, .
\end{equation}
Hence, substitution into (\ref{cum})  yields 
\begin{equation} \label{K_RD}
{\bf K}=
\left(
\begin{array}{ccc}
         0 &  0                  &  2            \cr
    \Delta & -2\alpha_0 +2\Delta & -2 \kappa   \cr
   -\kappa &  1                  &   -\alpha_0 +\Delta/2
\end{array}
\right) \, .
\end{equation}
From the eigenvalues of $\mathbf{K}$ we obtain $\lambda^\star$
following Eq.~(\ref{lambdastar}).
A theoretical expression for $\lambda$ can be found in 
Ref.~\cite{leprovost}.
Fig.~\ref{fig:RD0} exhibits numerical and analytical results
for both exponents, as a function of the noise intensity.  

Note that in the cases considered above 
$\lambda$ and $\lambda^\star$
do not coincide. 
We have checked that the difference between them (which is a
quantifier of the degree of intermittency of the dynamics) may be
controlled by suitable choice of the parameters of the
oscillator.
We preferred to consider intermittent cases, because it is
in these regimes that the numerical difficulties arise, as we
discuss in the following section.
We remark that there are situations of interest where both
exponents practically coincide, e.g., for a dilute Lennard-Jones
gas \cite{av4}.
In such cases a theory capable of estimating $\lambda^\star$ will
also produce a good estimate for the standard Lyapunov exponent
$\lambda$.

\section{Numerical treatment} \label{sec:diff}

\begin{figure}[h] 
\begin{center}
\includegraphics*[bb=110 420 510 690, width=0.45\textwidth]{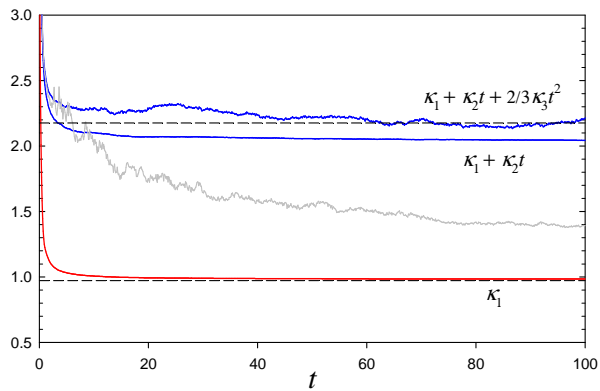}
\end{center}
\caption{  \label{fig:diff} 
Numerical difficulties in the calculation of $\lambda^\star$.
We plot the finite-time exponent $\lambda(t)\equiv \kappa_1(t)$ as 
a function of time (red line) for the random frequency oscillator 
with $\alpha=0$ and $\Delta =50$. 
Averages were computed over $10^5$ realizations.    
Also plotted are the corrections arising from the second
and third cumulants of the distribution of finite-time
Lyapunov exponents Eq.~(\ref{local}) (blue). 
For comparison we also show the straightforward average 
$\ln \langle | \eta (t; x_0,\eta_0)|^2 \rangle/2t$ 
(light gray). 
Dashed lines correspond to the theoretical asymptotic values. 
}
\end{figure}

Numerical simulations were performed by means of the Euler 
algorithm with time step $dt=10^{-3}$. 
For each trajectory we computed the norm 
$|\eta(t)|^2=q^2+p^2$ 
as a function of time $t$. 
The Lyapunov exponent is approximated by the average
over initial conditions of the finite-time exponents:
\begin{equation}
\lambda \approx 
\frac{1}{t} 
\langle 
\ln | \eta (t; x_0,\eta_0)| 
\rangle 
\equiv
\langle 
\lambda(t;x_0,\eta_0)
\rangle 
\;,
\end{equation}
where $t$ is large enough to guarantee the convergence of the
average to the desired precision.
In order to obtain the asymptotic value of the generalized
exponent (\ref{liapstarav}),
in principle, one must calculate the squared-norm averaged over 
several realizations at a given large time. 
However, we must keep in mind that such an average is dominated 
by the extreme positive values of the local exponent $\lambda(t)$. 
Hence, direct averaging over $|\eta(t)|^2$ 
may yield spurious results whenever the variance of $\lambda(t)$ 
fails to vanish with time fast enough. 
To avoid this problem, instead of the straightforward averaging,
we preferred to estimate the local generalized exponent from
the cumulants of the distribution of $\lambda(t)$:
\begin{equation} \label{local}
\lambda^\star(t)=
\frac{\ln\langle | \eta (t)|^2  \rangle}{2t} 
= 
\frac{\ln\langle {\rm e}^{2\lambda(t)t} \rangle}{2t} 
=
\sum_{n\ge 1} \frac{(2t)^{n-1}}{n!} \, \kappa_n(t)\,,
\end{equation}
where $\kappa_n$ are the $n$th-order cumulants of the 
distribution of local exponents $\lambda(t)$.
Fig.~\ref{fig:diff} illustrates, for the white-noise 
random frequency oscillator, 
the $\langle \lambda(t) \rangle $ as a function of time 
(first-order truncation of (\ref{local})), 
as well as the expansion (\ref{local}) truncated at 
the second and third orders. 
For comparison, also plotted is the crude estimate 
(\ref{liapstarav}).  
Clearly, the expansion (\ref{local}) has to be considered in 
order to properly estimate $\lambda^\star$. 
In Figs.~\ref{fig:RF0} and \ref{fig:RD0}, $\lambda^\star$ was 
numerically computed from the third-order truncation, 
because the next (noisier) terms do not contribute significantly.

\section{Correlated noise}

For white noise fluctuations, either in the frequency or in 
the damping, we have verified in Sec.~3 
(see Figs.~\ref{fig:RF0} and \ref{fig:RD0}), 
that our theory for $\lambda^\star$ is in agreement with numerical 
results, 
provided the later are obtained by means of the procedure 
described in the preceding section. 
The analysis in Sec.~3 also allows to quantify the discrepancy 
between $\lambda$ and $\lambda^\star$, 
which typically increases with increasing amplitude of 
the fluctuations. 

Now we shall analyze the effect of introducing noise 
correlations. 
We consider again the case of a random frequency, 
as in Eq.~(\ref{kappa}), 
but now the noise is a zero-mean Ornstein-Ulhenbeck 
process, i.e., with correlation function
\begin{equation}
\langle\xi(t)\xi(t')\rangle=
\frac{\Delta}{2\tau}\exp(-|t-t'|/\tau) \equiv
\sigma^2 \exp(-|t-t'|/\tau)
\, .
\end{equation}
For simplicity we set $\alpha=0$ and $\kappa_0=0$. 
By inserting Eq.~(\ref{B_RF}) into 
Eq.~(\ref{cum}), the second-cumulant matrix 
$\mathbf{K}^{(2)}$ becomes 
\begin{equation}  \label{K_2}
\mathbf{K}^{(2)}=\left(
\begin{array}{ccc}
               0 &                  0 &  2                 \cr
          \Delta & -2\Delta \, \tau^2 &  0                 \cr
  \Delta \, \tau &                  1 & -2 \Delta \, \tau^2 
\end{array}
\right) \,.
\end{equation}
Notice that in the limit $\tau\to 0$ the white-noise case 
is recovered. 

\begin{figure}[h] 
\begin{center}
\includegraphics*[bb=110 420 520 690, width=0.45\textwidth]{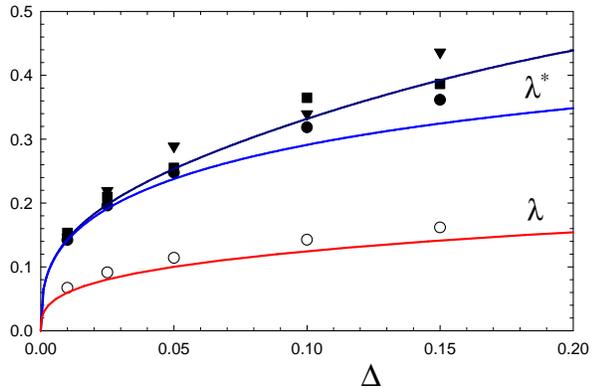}
\end{center}
\caption{\label{fig:RFOU}  Harmonic oscillator with correlated random frequency.
We show $\lambda$ and $\lambda^\star$ as a function of the noise amplitude $\Delta$.
Solid lines correspond to the theoretical results obtained from 
(\ref{lambdastar},\ref{K_2}) (blue) and (\ref{lambdastar},\ref{K_4})  (dark blue) 
for $\lambda^\star$ and to the approximate expression (using the decoupling ansatz) 
following Ref.~\cite{mallick} for $\lambda$.
Symbols are the results of numerical simulations 
(averaged over $10^5$ trajectories), corresponding to the 
second (circles), third (squares) and fourth (triangles) order corrections of (\ref{local}). 
Parameters are $\alpha=0$, $\kappa_0=0$ and $\tau=1$. 
}
\end{figure}

In the presence of correlations the second-order truncation 
of the cumulant expansion (\ref{cum}) is not exact.
In order to improve the theory one must calculate higher
cumulants. For the present case the third cumulant is null.
Explicit expressions for the fourth cumulant were given 
by Fox \cite{fox} and by Breuer et al  \cite{breuer}.
So, the fourth cumulant can be calculated without great effort
(with the aid of algebraic manipulation programs).  
The fourth order approximation to $\mathbf{K}$ reads
\begin{equation}  \label{K_4}
\mathbf{K}^{(4)}=\mathbf{K}^{(2)}+ \frac{1}{2}\Delta^2\tau^3
\left(
\begin{array}{ccc}
     0 &         0 &         0  \cr
    13 &  74\tau^2 &   -57\tau  \cr
17\tau & 173\tau^3 & -99\tau^2 
\end{array}
\right) \,.
\end{equation}
The comparison between the theoretical results for $\lambda^\star$ 
(with the second (blue) and fourth (dark blue) order corrections) 
and numerical outcomes is shown in Fig.~\ref{fig:RFOU}. 
Notice that in numerical estimates the fourth-order correction is 
very small in comparison with the third-order one, suggesting that
the cumulant expansion is rapidly converging.

\subsection{Kubo number}

The perturbation parameter controlling the convergence of the
cumulant expansion is the so-called Kubo number $\varepsilon$. 
General considerations led van Kampen \cite{vankampen} to
conclude that the Kubo number is the product of the amplitude
of the fluctuations and the correlation time, that is
$\sigma \tau$.
However, in the present case it is clear that such a 
combination is not adimensional. 
The correct Kubo number is instead 
\begin{equation}
\varepsilon = 
\sigma \tau^2  = 
\sqrt{\frac{\Delta \tau^3}{2}} \, . 
\end{equation}
This can be checked explicitly from the second and fourth
cumulants above.
Consider, for instance, the element ${\bf K}_{21}$, which 
dominates the Lyapunov exponent for small correlation
times:
\begin{equation}
{\bf K}_{21} = 
\Delta + 
\frac{13}{2}\Delta^2\tau^3 + \dots =
\Delta \left(1+ \frac{13}{2}\Delta\tau^3 + \dots \right) \, . 
\end{equation}
In the white-noise limit, i.e., $\tau \to 0$ with $\Delta$
fixed, the Kubo number tends to zero --as it should be.

\section{Final remarks}

We have taken the first step towards the 
application 
of the cumulant expansion to calculate the largest 
Lyapunov exponent of a dilute gas.

The case of white-noise fluctuations
(either in the frequency or in the damping)
was considered first.
This study was very useful to understand the difficulties 
behind the numerical calculation of the generalized
exponent $\lambda^\star$. 
It was verified that $\lambda^\star$ can be obtained with
a satisfactory precision by using the cumulant expansion 
for the distribution of the finite-time Lyapunov exponent.
 
We also analyzed briefly the case of correlated noise.
For the Ornstein-Uhlenbeck noise we were able to obtain
the fourth cumulant contribution to the analytical 
$\lambda^\star$, which showed an improvement with respect 
to the second order truncation, when compared with numerical 
outcomes.  
Moreover, we showed that the correct perturbative parameter for
the present problem is the product $\sigma \tau^2$,
and not $\sigma \tau$, as a literal reading of
van Kampen's discussion \cite{vankampen} would
suggest.

It is expected that the present results will be helpful for the 
correct application of the cumulant approach
in higher dimensionality systems,
as well as 
for the numerical checking of its validity.

\section*{Acknowledgements:}

We acknowledge Brazilian agencies Faperj and CNPq for partial 
financial support. \\[5mm]


\begin{thebibliography}{99}


 
\bibitem{krylov}  
Krylov NS 1979 
{\em Works on the Foundations of Statistical Physics} 
(Princeton University Press, Princeton)

\bibitem{ma}
Ma S-K 1985
{\em Statistical Mechanics}
(World Scientific, Singapore)

\bibitem{sinai}  
Sinai YaG 1970  
{\em Russ. Math. Surv.} {\bf 25} 137

\bibitem{vanb1} 
van Beijeren H, Dorfman JR 1995
{\em Phys. Rev. Lett.} {\bf 74}  4412

\bibitem{vanb2} 
van Beijeren H, Latz A, Dorfman JR 1998
{\em Phys. Rev. E} {\bf 57}  4077

\bibitem{vanzon} 
van Zon R, van Beijeren H, Dellago Ch 1998
{\em Phys. Rev. Lett.} {\bf 80}  2035

\bibitem{kruis} 
Kruis HV, Panja D, van Beijeren H 2006 
{\em J. Stat. Phys.} {\bf 124} 823

\bibitem{dorfman}
Dorfman JR 1999
{\em An Introduction to Chaos in Nonequilibrium
Statistical Mechanics}
(Cambridge University Press, Cambridge, UK)

\bibitem{dellago1} 
Dellago Ch, Posch HA, Hoover WG 1996 
{\em Phys. Rev. E} {\bf 53} 1485

\bibitem{dellago2} 
Dellago Ch, Posch HA 1997 
{\em Physica A} {\bf 240} 68

\bibitem{kimball} 
Kimball JC 2001 
{\em Phys. Rev. E} {\bf 63} 066216

\bibitem{elyutin} 
Elyutin PV 2004 
{\em Phys. Lett. A} {\bf 331} 153

\bibitem{benettin}  
Benettin G, Galgani L, Strelcyn J-M 1976
{\em Phys. Rev. A} {\bf 14} 2338

\bibitem{vankampen} 
van Kampen NG 1981 
\emph{Stochastic Processes in Physics and Chemistry} 
(North-Holland, Amsterdam)

\bibitem{kubo}
Kubo R 1962
{\em J. Phys. Soc. Japan} {\bf 17} 1100

\bibitem{fox}
Fox RF 1974
{\em J. Math. Phys} {\bf 15} 1479

\bibitem{benzi}
Benzi R, Paladin G, Parisi G, Vulpiani A 1985 
{\em J. Phys. A} {\bf 18} 2157

\bibitem{castiglione}
Castiglione P, Falcioni M, Lesne A, Vulpiani A 2008
{\em Chaos and Coarse Graining in Statistical Mechanics}
(Cambridge University Press, New York, 2008)

\bibitem{barnett1}
Barnett DM, Tajima T, Nishihara K, Ueshima Y, Furukawa H 1996 
{\em Phys. Rev. Lett.} {\bf 76} 1812

\bibitem{barnett2}
Barnett DM, Tajima T, Nishihara K, Ueshima Y, Furukawa H 1997 
{\em Phys. Rev. E} {\bf 55} 3439

\bibitem{barnett3}
Barnett DM, Tajima T 1996 
{\em Phys. Rev. E} {\bf 54} 6084

\bibitem{pettini1}
Casetti L, Livi R, Pettini M 1995 
{\em Phys. Rev. Lett.} {\bf 74} 375

\bibitem{pettini2}
Casetti L, Clementi C, Pettini M, 1996 
{\em Phys. Rev. E} {\bf 54} 5969

\bibitem{pettini3} 
Casetti L, Pettini M, Cohen EGD. 2000 
{\em Phys. Rep.} {\bf 337} 238

\bibitem{av1}
Vallejos RO, Anteneodo C 2002 
{\em Phys. Rev. E} {\bf 66} 021110  

\bibitem{av2}
Anteneodo C, Maia RNP,  Vallejos RO, 2003 
{\em Phys. Rev. E} {\bf 68} 036120   

\bibitem{av3}
Vallejos RO, Anteneodo C 2004 
{\em Physica A} {\bf 340} 178 

\bibitem{comment}
Torcini A, Dellago Ch, Posch HA 1999 
{\em Phys. Rev. Lett.} {\bf 83} 2676; Barnett DM, Tajima T, Ueshima Y 1999
{\em ibid.} {\bf 83} 2677

\bibitem{av4}
Anteneodo C, Cirto L, Vallejos RO (unpublished)


\bibitem{mallick} 
Mallick K, Peyneau PE 2006 
{\em Physica D} {\bf 221} 72

\bibitem{leprovost} 
Leprovost N, Auma\^itre S, Mallick K 2006 
{\em Eur. Phys. J. B} {\bf 49} 453

\bibitem{breuer}
Breuer H-P, Ma A, Petruccione F 2002
{\em preprint} arXiv:quant-ph/0209153v1


\bibitem{risken} 
Risken H 1984 
\emph{The Fokker-Planck Equation: 
Methods of Solution and Applications} 
(Springer-Verlag, Berlin)

\bibitem{gardiner} 
Gardiner C\ W 1985 
{\em Handbook of stochastic methods for
Physics, Chemistry and Natural Sciences} 
(Springer-Verlag, Berlin)
 
\end{thebibliography}
\end{document}